\documentclass[12pt]{article}
\usepackage[pdftex,bookmarksopen=true,bookmarks=true,unicode,setpagesize]{hyperref}
\hypersetup{colorlinks=true,linkcolor=black,citecolor=black}

\usepackage{amssymb, amsmath, amsthm}
\usepackage{mathrsfs}

\usepackage{xcolor}

\usepackage{mathtools}

\allowdisplaybreaks
\usepackage{cite}

\newcommand\R{\mathbb{R}}
\newcommand\N{\mathbb{N}}

\newcommand{\vertiii}[1]{{\vert\kern-0.25ex\vert\kern-0.25ex\vert #1
    \vert\kern-0.25ex\vert\kern-0.25ex\vert}}

\newtheorem{theorem}{Theorem}

\theoremstyle{remark}

\theoremstyle{remark}

\theoremstyle{remark}
\newtheorem{remark}[theorem]{Remark}

\begin{document}

\vspace{-20mm}
\begin{center}{\Large \bf 
The projection spectral theorem, quasi-free states and point processes}
\end{center}

\begin{center}
{\it Dedicated to the 100th anniversary of Yuri Makarovych Berezansky's birth}
\end{center}

{\large Eugene Lytvynov}\\ Department of Mathematics, Swansea University, Bay Campus,  Swansea SA1 8EN, U.K.;
e-mail: \texttt{e.lytvynov@swansea.ac.uk}\vspace{2mm}

{\small
\begin{center}
{\bf Abstract}
 \end{center}

\noindent  In this review paper, we demonstrate that several classes of point processes in a locally compact Polish space $X$ appear as the joint spectral measure of a rigorously defined particle density of a representation of the canonical anticommutation relations (CAR) or the canonical commutation relations (CCR) in a Fock space. For these representations of the CAR/CCR, the vacuum state on the corresponding $*$-algebra is quasi-free. The classes of point process that arise in such a way include determinantal and permanental point processes.

 } \vspace{2mm}


 \vspace{2mm}


\section{Introduction}

The projection spectral theorem for a family of commuting self-adjoint (or more generally, normal) operators and the corresponding Fourier transform in generalized joint eigenvectors of this family play a fundamental role in functional analysis and its applications, in particular, to infinite dimensional analysis and mathematical physics. In the case where the operator family has additionally a cyclic vector, this result allows one to realize the operators from the family as multiplication operators   in an $L^2$-space with respect to a probability measure. This theory was studied by many authors, including Yu.M.~Berezansky, L.~G\r{a}rding, I.M.~Gelfand, A.G.~Kostyuchechko, G.~Kats, and K.~Maurin. In the most general setting, this result was proved by Berezansky in \cite{ber_1984}, see also \cite{ber_1985,ber_1978} and Chapter 3 in \cite{BK}.
In particular, the theory of Berezansky allowed one to consider families of an uncountable number of operators

The projection spectral theorem below (Theorem~\ref{spectral}) follows immediately from the results of \cite[Chapter 3]{BK}. It is presented here in the form as in our review paper \cite{L1}. But first, let us remind the reader that we say that 
$\Phi \subset H \subset \Phi'$ is a standard triple if  $H$ is a separable Hilbert space, $\Phi$ is a nuclear space that is densely and continuously embedded into $H$, and $\Phi'$ is the dual space of $\Phi$ with respect to the center space $H$. The latter means that the dual pairing between $\omega\in\Phi'$ and $\varphi\in\Phi$, denoted by $\langle \omega,\varphi\rangle$, is determined by the  inner product in $H$. Note that a standard triple can be real (in which case all three spaces in the triple are real) or complex (in which case all three spaces in the triple are complex, and one also typically assumes that the complex Hilbert space $H$ is the complexification of a real Hilbert space $H_\mathbb R$).

\begin{theorem} \label{spectral}
Let $\Phi \subset H \subset \Phi'$ be a real standard triple, and let $\Psi \subset \mathcal{F} \subset \Psi' $ be a complex standard triple. Assume  we are given a family $(A(\varphi))_{\varphi \in \Phi} $ of Hermitian operators in $\mathcal{F}$ such that
\begin{enumerate}
\item $D(A(\varphi )) = \Psi$, $\varphi \in\Phi $;

\item $A(\varphi )\Psi \subset \Psi$ for each $\varphi \in \Phi$,
and furthermore $ A(\varphi ):\Psi \rightarrow \Psi $ is continuous;

\item $A(\varphi_1)A(\varphi_2)f = A(\varphi_2)A(\varphi_1)f$ for each   $\varphi_1,\varphi_2\in\Phi$ and $f \in \Psi$;

\item for all $f, g \in \Psi$, the mapping  $ \Phi \ni \varphi \mapsto (A(\varphi )f,g)_{\mathcal{F}} \in \mathbb{C}$  is continuous;

\item there exists a vector $\Omega$ in $\Psi$ which is cyclic for $(A(\varphi ))_{\varphi \in \Phi }$, i.e., the linear span of the set
$$\{\Omega,\, A(\varphi_1) \cdots A(\varphi_k)\Omega \mid \varphi_1, \dots, \varphi_k \in \Phi , \ k \in \mathbb{N}\}$$ is dense in $\mathcal {F}$;

\item for any $f \in \Psi $ and  $\varphi \in \Phi $, the vector $f$ is analytic for the operator $A(\varphi )$, i.e., for some $t>0$,
$$\sum_{n=0}^\infty t^n\,\frac{\|(A(\varphi)^nf\|_{\mathcal F}}{n!}<\infty.$$
\end{enumerate}

Then, each operator $A(\varphi )$, $\varphi \in\Phi $, is essentially self-adjoint in $\mathcal F$, and we denote its closure by $(\tilde{A}(\varphi ),D(\tilde{A}(\varphi )))$.
The operators $(\tilde A(\varphi))_{\varphi\in\Phi}$ commute in the sense of commutation of their projection-valued measures (resolutions of the identity). Furthermore, there exists a unique probability measure
$\mu$ on $(\Phi',\mathcal{C}(\Phi'))$ such that the linear operator $I: \mathcal{F} \rightarrow L^2(\Phi',\mu)$ satisfying  
$I \Omega = 1$ and
\begin{align*}
I(\tilde{A}(\varphi_1) \cdots \tilde{A}(\varphi_n)\Omega) &= I(A(\varphi_1) \cdots A(\varphi_n)\Omega)\\&= \langle  \cdot, \varphi_1 \rangle \cdots \langle \cdot ,\varphi_n \rangle
\in L^2(\Phi', \mu)
\end{align*}
is unitary. Here, $\mathcal C(\Phi')$ denotes the cylinder $\sigma$-algebra on $\Phi'$.  

Under the action of $I$, each  operator $(\tilde{A}(\varphi),D(\tilde{A}(\varphi)))$, $\varphi\in\Phi$, becomes the operator of multiplication by
$\langle \cdot ,\varphi \rangle$ in $L^2(\Phi', \mu)$, denoted by $M(\varphi)$, i.e., 
$$ D(M(\varphi))=\big\{F \in L^2(\Phi', \mu)\mid \langle\cdot, \varphi\rangle F \in L^2(\Phi', \mu)\big\}$$
and for each $F\in  D(M(\varphi))$,
$$ (M(\varphi) F)(\omega)=\langle\omega,\varphi\rangle F(\omega).$$
\end{theorem}

The probability measure $\mu$ on $\Phi'$ from Theorem~\ref{spectral} is called the joint spectral measure of the family of (commuting self-adjoint) operators $(\tilde A(\varphi))_{\varphi\in\Phi}$ (with respect to the cyclic vector $\Omega$).

Berezansky et al.\ \cite{BKKL} applied the projection spectral theorem to study of point processes and their correlation measures. The paper \cite{BKKL} was  influenced by Kondratiev and Kuna's studies \cite{KK} of harmonic analysis on configuration spaces.  

Let $X$ be a locally compact Polish space, let $\Gamma(X)$ denote the set of simple configurations in $X$, and equip $\Gamma(X)$ with the cylinder $\sigma$-algebra $\mathcal C(\Gamma(X))$. Let $\Gamma_0(X)$ denote the (measurable) subset of $\Gamma(X)$ that consists of all finite configurations in $X$. A point process  in $X$ is a probability measure on $\Gamma(X)$. To each point process $\mu$ in $X$, there corresponds its correlation measure $\theta=\theta_\mu$, which is a (generally speaking infinite) measure on $\Gamma_0(X)$. The measure $\theta$ can be defined through the $K$-transform, which maps functions on $\Gamma_0(X)$ to functions on $\Gamma(X)$. In paper \cite{KK}, the $K$-transform was studied  as an operator $K:L^1(\Gamma_0(X),\theta)\to L^1(\Gamma(X),\mu)$ that satisfies\footnote{In fact, for a given point process $\mu$, formula \eqref{vydtde6e6u} uniquely determines its correlation measure $\theta$.}
\begin{equation}\label{vydtde6e6u}
\int_{\Gamma_0(X)}G(\eta)\,\theta(d\eta)=\int_{\Gamma(X)} (KG)(\gamma)\,\mu(d\gamma),\quad G\in L^1(\Gamma_0(X),\theta).\end{equation}
Furthermore, a convolution $\star$ of functions on $\Gamma_0(X)$ was explicitly constructed in \cite{KK}, and this convolution  satisfies $(K(G_1\star G_2))(\gamma)=(KG_1)(\gamma)(KG_2)(\gamma)$. 

Let  $\theta$ be the correlation measure of a point process $\mu$. Define
\begin{equation}\label{vdyrt6y}
\mathcal R_\theta=\{G\in L^1(\Gamma_0(X),\theta): KG\in L^2(\Gamma(X),\mu)\}.
\end{equation}
The measure $\theta$ is $\star$-positive definite in the following sense:
\begin{equation}\label{dydydy}
\int_{\Gamma_0(X)}(G\star\overline G)(\eta)\,\theta(d\eta)\ge0\quad \text{for each }G\in\mathcal R_\theta.
\end{equation}
Hence, $\mathcal R_\theta$ can be naturally interpreted as a complex Hilbert space with the inner product 
\begin{equation}\label{vgcydy}
(G_1,G_2)_{\mathcal R_\theta}=\int_{\Gamma_0(X)}(G_1\star \overline{G_2})(\eta)\,\theta(d\eta),\quad G_1,G_2\in\mathcal R_\theta.
\end{equation}

Now assume that $X$ is a Riemannian manifold  and $\theta$ is a measure on $\Gamma_0(X)$ that is $\star$-positive definite, and satisfies some additional technical assumptions.  We define a complex Hilbert space $\mathcal R_\theta$ of functions on $\Gamma_0(X)$, with the inner product given by \eqref{vgcydy}. It is not difficult to construct  a standard triple $\Psi\subset\mathcal R_\theta\subset\Psi'$.

Let $\mathcal D(X)$ denote the nuclear space of all smooth functions $\varphi:X\to\mathbb R$ with compact support. For each $\varphi\in\mathcal D(X)$, one can define a Hermitian operator  $A(\varphi)$ in $\mathcal R_\theta$, with  domain $\Psi$, that acts as follows: $A(\varphi)G=\varphi\star G $. 
In this formula, $\varphi\in\mathcal D(X)$ is naturally identified with a function on $\Gamma_0(X)$ (see Section~\ref{vgftyfdytdty} for details). 
 It was proved in \cite{BKKL} that
  the family of operators $(A(\varphi))_{\varphi\in\mathcal D(X)}$ satisfies the conditions of Theorem~\ref{spectral}. The corresponding spectral measure $\mu$, defined initially on $\mathcal D'(X)$, was shown to be concentrated on the configuration space $\Gamma(X)$, and furthermore, $\mu$ is nothing but the point process for which $\theta$ is its correlation  measure. Furthermore the unitary operator $I:\mathcal R_\theta\to L^2(\Gamma(X),\mu)$ from Theorem~\ref{spectral} is the restriction of the $K$-transform $K:L^1(\Gamma_0(X),\theta)\to L^1(\Gamma(X),\mu)$ to $\mathcal R_\theta\subset L^1(\Gamma_0(X),\theta)$. 
 
The studies of the $K$-transform as the joint spectral measure of a family of commuting self-adjoint operators  were 
advanced  in \cite{LM}. That paper dealt with the more general case of a locally compact Polish space $X$,  and it treated a family of commuting Hermitian operators $\rho(\Delta)$, indexed by all pre-compact Borel subsets $\Delta$ of $X$. For such a family of Hermitian operators, the notion of its correlation measure $\theta$ on $\Gamma_0(X)$ was given. Note that, unlike the case of a point process, the correlation measure of a family of Hermitian operators does not need to necessarily exist. However, if such a correlation measure $\theta$ does exist, then there also exists a point process $\mu$ in $X$ for which $\theta$ is its correlation measure, while $\mu$ is the joint spectral measure of the (closures of the) operators $\rho(\Delta)$.  

This result from \cite{LM} was applied to several families of Hermitian operators $\rho(\Delta)$ that arise as the particle density of a representation of the canonical anticommutation relations (CAR) or the canonical commutation relations (CCR). These representations are typically associated  with a quasi-free state on the corresponding CAR/CCR $*$-algebra. The point processes that were discussed via this  ansatz included Poisson point processes, determinantal and permanental point processes, and hafnian point processes. The aim of this paper is to briefly review these results.  

The paper is organized as follows. In Section~\ref{niguyfu7}, we will recall some key  definitions and results related to the $K$-transform. In Section~\ref{vgftyfdytdty}, we will describe the main result of paper \cite{BKKL}, while in Section~\ref{jgbiguyifguy} we will recall the above mentioned result from \cite{LM}. In Section~\ref{vgcgcddt}, we will recall basic definitions and results related to  quasi-free  and  gauge-invariant  states on the CAR and CCR $*$-algebras. In Section~\ref{fxdsdese}, we will discuss symmetric and antisymmetric Fock spaces,  creation and annihilation operators in them, and the differential second quantization.  Finally, in Section~\ref{bfydr7er74r}, we will review several applications of the theorem from \cite{LM} to particle densities and point processes. Our key references  in Section~\ref{bfydr7er74r} are \cite{LM,L2,AL1,AL2}.

\section{Point process and its correlation measure}\label{niguyfu7}

We mostly present in this section results from \cite{KK} (see also the references therein), sometimes in a slightly modified form.   

Let $X$ be a locally compact Polish space, let $\mathcal B(X)$ denote the Borel $\sigma$-algebra on $X$, and let $\mathcal B_0(X)$ denote the algebra of all pre-compact sets from $\mathcal B(X)$. Let $\sigma$ be a measure on $(X,\mathcal B(X))$ which is non-atomic (i.e., $\sigma(\{x\})=0$ for all $x\in X$) and Radon (i.e., $\sigma(\Delta)<\infty$ for all $\Delta\in\mathcal B_0(X)$). For applications, the most important example is $X=\R^d$ and $\sigma(dx)=dx$ is the Lebesgue measure.

A  (simple) configuration $\gamma$ in $X$ is a Radon measure on $X$ of the form $\gamma=\sum_i\delta_{x_i}$, where $\delta_{x}$ denotes the Dirac measure with mass at $x$, and $x_i\ne x_j$ if $i\ne j$. By definition, the zero measure on $X$ is a  configuration. 
Note that, since $\gamma$ is a Radon measure, it has a finite number of atoms $x_i$ in each compact set in $X$, however the total mass of $X$ can be $\infty$. Let $\Gamma(X)$ denote the set of all configurations $\gamma$ in $X$. Let $\mathcal C(\Gamma(X))$ denote the minimal $\sigma$-algebra on $\Gamma(X)$ with respect to which the  map $\Gamma(X)\ni\gamma\mapsto\gamma(\Delta)$ is measurable for each $\Delta\in\mathcal B_0(X)$. The $\sigma$-algebra  $\mathcal C(\Gamma(X))$ coincides with the Borel $\sigma$-algebra of the vague topology on $\Gamma_X$, i.e., it is the minimal $\sigma$-algebra on $\Gamma_X$ with respect to which each map of the form  $\Gamma_X
\ni\gamma\mapsto \langle\gamma,\varphi\rangle=\int_X\varphi\,d\gamma \in\mathbb R$ is continuous, where  $\varphi:X\to\mathbb R$ is a continuous function with compact support.

A  (simple) point process in $X$ is a probability measure on $(\Gamma(X),\mathcal C(\Gamma(X)))$.

 A measure on $X^n$ is called symmetric if it remains invariant under the natural action of the symmetric group $\mathfrak S_n$ on $X^n$. 
 For each $\gamma=\sum_i\delta_{x_i}\in\Gamma(X)$, the  spatial falling factorial $(\gamma)_n$ is the symmetric measure on $X^n$  of the form
\begin{equation*}
(\gamma)_n=\sum_{i_1}\sum_{i_2\ne i_1}\dotsm\sum_{i_n\ne i_1,\dots, i_n\ne i_{n-1}}\delta_{(x_{i_1}, x_{i_2}, \dots,x_{i_n})}.\end{equation*}
The measures $(\gamma)_n$ ($n\in\mathbb N$) satisfy the recurrence formula  
\begin{align}
&(\gamma)_1(\Delta)=\gamma(\Delta),\notag\\
&(\gamma)_{n+1}(\Delta_1\times\dots\times\Delta_{n+1})=\gamma(\Delta_{n+1})(\gamma)_{n}(\Delta_1\times\dots\times\Delta_{n})
\notag\\
&\qquad-\sum_{i=1}^n
(\gamma)_n(\Delta_1\times\Delta_{i-1}\times(\Delta_i\cap\Delta_{n+1})\times\Delta_{i+1}\times\dots\times\Delta_n),\quad n\in\mathbb N,
\label{gfxtdsxzetrsresa8}
 \end{align}
where $\Delta,\Delta_1,\dots,\Delta_{n+1}\in\mathcal B_0(X)$. For $n\ge2$, denote
$$X^{(n)}=\{(x_1,\dots,x_n)\in X^n: x_i\ne x_j\text{ if }i\ne j\}.$$
Let also $X^{(1)}=X$. For $\gamma\in\Gamma(X)$ and $n\in\mathbb N$, the measure $(\gamma)_n$ is concentrated on $X^{(n)}$, i.e., $(\gamma)_n(X^n\setminus X^{(n)})=0$.

Let $\mu$ be a point process in $X$. For $n\in\mathbb N$, the  $n$th correlation measure of $\mu$ is the symmetric measure $\theta^{(n)}$ on $X^n$ defined by
\begin{equation*}
\theta^{(n)}(dx_1\dotsm dx_n)=\frac1{n!}\int_{\Gamma(X)}(\gamma)_n(dx_1\dotsm dx_n)\,\mu(d\gamma).
\end{equation*}
The measure $\theta^{(n)}$ is concentrated on $X^{(n)}$. 
If each measure $\theta^{(n)}$ is absolutely continuous with respect to $\sigma^{\otimes n}$, then the symmetric functions $k^{(n)}:X^{n}\to[0,\infty)$ satisfying 
\begin{equation}\label{5w738}
d\theta^{(n)}=\frac1{n!}\,k^{(n)}d\sigma^{\otimes n}\end{equation} 
are called the  correlation functions of the point process $\mu$. Under a very weak assumption,  the correlation measures  (or the correlations functions) uniquely identify a point process, see  \cite{Lenard}.

Let $\Gamma_0(X)=\{\gamma\in\Gamma(X): \gamma(X)<\infty\}$.    Then $\Gamma_0(X)=\bigcup_{n=0}^\infty\Gamma^{(n)}(X)$, where 
$$\Gamma^{(n)}(X)=\{\gamma\in \Gamma_0(X): \gamma(X)=n\}.$$ 
For each $n\in\mathbb N$, consider the map 
\begin{equation}\label{cfxdszrts}
X^{(n)}\ni(x_1,\dots,x_n)\mapsto\sum_{i=1}^n\delta_{x_i}\in\Gamma^{(n)}(X).
\end{equation}
 Let $\mathcal B(\Gamma^{(n)}(X))$ denote the $\sigma$-algebra on $\Gamma^{(n)}(X)$ that is the image of the Borel $\sigma$-algebra $\mathcal B(X^{(n)})$ under the map in \eqref{cfxdszrts}. (Equivalently, instead of  $\mathcal B(X^{(n)})$, one can use the sub-$\sigma$-algebra  $\mathcal B_{\mathrm{sym}}(X^{(n)})$ that consists of all symmetric sets from $\mathcal B(X^{(n)})$). Let $\mathcal B(\Gamma_0(X))$ denote the $\sigma$-algebra on $\Gamma_0(X)$  whose trace on each $\Gamma^{(n)}(X)$ ($n\in\mathbb N$) coincides with $\mathcal B(\Gamma^{(n)}(X))$.  In fact, $\mathcal B(\Gamma_0(X))$ coincides with the trace of the $\sigma$-algebra $\mathcal C(\Gamma(X)))$ on $\Gamma_0(X)$.

Let $\mu$ be a point process and let $(\theta^{(n)})_{n=1}^\infty$ be the sequence of its correlation measures. It is convenient to interpret the sequence $(\theta^{(n)})_{n=1}^\infty$ as a single measure $\theta$ on $(\Gamma_0(X),\mathcal B(\Gamma_0(X)))$. To this end, let us keep the notation $\theta^{(n)}$ for the push-forward of the measure $\theta^{(n)}$ on $X^{(n)}$ under  the map in \eqref{cfxdszrts}. Then, we define the measure $\theta$ on $(\Gamma_0(X),\mathcal B(\Gamma_0(X)))$ that  coincides with $\theta^{(n)}$ on each $\Gamma^{(n)}(X)$ for $n\in\mathbb N$, and $\theta(\Gamma^{(0)}(X))=1$. The measure $\theta$ is called the correlation measure (on $\Gamma_0(X)$) of the point process $\mu$. 

The correlation measure $\theta$ can also be defined through the so-called $K$-transform. For a configuration $\gamma\in\Gamma(X)$ and a finite configuration $\eta\in \Gamma_0(X)$, we will write $\eta\prec\gamma$ if $\gamma-\eta\in\Gamma(X)$, i.e., the support of $\eta$ is a subset of the support of $\gamma$.  For a measurable function $G:\Gamma_0(X)\to[0,\infty)$, we define  a measurable function $KG:\Gamma(X)\to[0,\infty]$ by  
$$(KG)(\gamma)=\sum_{\eta\in\Gamma_0(X),\ \eta\prec\gamma}G(\eta),\quad \gamma\in\Gamma(X).$$ 
Then, for a point process $\mu$, its correlation measure $\theta$ on $\Gamma_0(X)$ satisfies
\begin{equation}\label{cfxdtstst6}
\theta(\Lambda)=\int_{\Gamma(X)} (K\chi_\Lambda)(\gamma)\,\mu(d\gamma),\quad \Lambda\in\mathcal B(\Gamma_0(X)),
\end{equation}
where $\chi_\Lambda$ is the indicator function of the set $\Lambda$. Formula \eqref{cfxdtstst6} serves as an equivalent definition of the correlation measure $\theta$.

The $K$-transform can be extended to act on a class of measurable functions\linebreak  $G:\Gamma_0(X)\to\mathbb R$. Indeed, let $G^+(\eta)=\max\{G(\eta),0\}$,  $G^-(\eta)=\max\{-G(\eta),0\}$, so that $G=G^+-G^-$. For $\eta\in\Gamma_0(X)$, we define $(KG)(\eta)=(KG^+)(\eta)-(KG^-)(\eta)$, provided both terms $(KG^+)(\eta)$ and $(KG^-)(\eta)$ are finite.  Similarly, one defines the $K$-transform of a function $G:\Gamma_0(X)\to\mathbb C$. 

If $\mu$ is a point process and $\theta$ is its correlation measure on $\Gamma_0(X)$, then, for each function $G\in L^1(\Gamma_0(X),\theta)$, the function $KG$ is well-defined $\mu$-a.e., $KG\in L^1(\Gamma(X),\mu)$, and furthermore formula \eqref{vydtde6e6u} holds. 

The $K$-transform leads to a natural convolution of functions on $\Gamma_0(X)$. For any measurable functions $G_1,G_2:\Gamma_0(X)\to\mathbb C$, we define a function $G_1\star G_2:\Gamma_0(X)\to\mathbb C$ by 
$$(G_1\star G_2)(\eta)=\sum_{\substack{\eta_1,\eta_2\in\Gamma_0(X)\\\eta_1\prec\eta,\ \eta_2\prec\eta}}G_1(\eta_1)G_2(\eta_2),\quad \eta\in\Gamma_0(X). $$
Then, 
\begin{equation}\label{vdrt6ue6}
(K(G_1\star G_2))(\gamma)=(KG_1)(\gamma)(KG_2)(\gamma)
\end{equation}
for each $\gamma\in\Gamma(X)$ such that both $(KG_1)(\gamma)$ and $(KG_2)(\gamma)$ are well defined and finite.

In view of formula \eqref{vdrt6ue6}, formula \eqref{dydydy} holds, where $\mathcal R_\theta$ is defined by \eqref{vdyrt6y}.

\section{Spectral representation of the correlation 
  measure}\label{vgftyfdytdty}

In this section, we will briefly explain the main result of Berezansky et al.\ \cite{BKKL}.

Let $X$ be a connected, oriented $C^\infty$ (non-compact) Riemannian manifold. Let $\sigma$ be the volume measure on $X$.

We will say that a function $G:\Gamma_0(X)\to\mathbb C$ has a bounded support if there exist a compact set $\Lambda\subset X$ and $N\in\mathbb N$ such that the function $G$ vanishes outside the set $\bigcup_{n=0}^N\Gamma^{(n)}(\Lambda)$, where $\Gamma^{(n)}(\Lambda)=\{\eta\in\Gamma^{(n)}(X):\eta(\Lambda^c)=0\}$. Let us denote by $\mathcal F(\Gamma_0(X))$ the set of all measurable bounded functions $G:\Gamma_0(X)\to\mathbb C$ with bounded support.  It is easy to see that, for each $G\in \mathcal F(\Gamma_0(X))$,  $(KG)(\gamma)$ is well-defined for all $\gamma\in \Gamma(X)$, and $KG$ is a bounded function on $\Gamma(X)$. 

Let $\theta$ be a measure on $\Gamma_0(X)$. We make the following assumptions about $\theta$:

\begin{description}

\item[{\rm (A1)}] {\it Normalization}: $\theta(\Gamma^{(0)}(X))=1$.

\item[{\rm (A2)}]
{\it Local finiteness}:
 For each $n\in\N$ and each compact set $\Lambda
\subset X$, we have $\theta(\Gamma_{\Lambda}^{(n)})<\infty$.

\item[{\rm (A3)}] {\it $\star$-positive definiteness}: For each $G\in\mathcal F(\Gamma_0(X))$, 
$\int_{\Gamma_0(X)}(G\star\overline G)(\eta)\,\theta(d\eta)\ge0$. 
\end{description}

We define an inner product on $\mathcal F(\Gamma_0(X))$ by
$$(G_1,G_2)_{\mathcal R_\theta}=\int_{\Gamma_0(X)} (G_1\star\overline{G_2})(\eta)\,\theta(d\eta).$$
Let $\mathcal R_\theta$ denote the complex Hilbert space obtained as the completion of $\mathcal F(\Gamma_0(X))$ in the norm induced by the inner product $(\cdot,\cdot)_{\mathcal R_\theta}$.  (As usual, one identifies in $\mathcal R_\theta $ any two functions $G_1,G_2\in \mathcal F(\Gamma_0(X))$ for which $\|G_1-G_2\|_{\mathcal R_\theta}=0$.)  

We denote by $\mathcal D(X)$ the nuclear space of all real-valued smooth functions on $X$ with compact support. Let $\mathcal D'(X)$ denote the dual space of $\mathcal D(X)$ with respect to the center space $L^2(X\to\mathbb R,\sigma)$. We denote by $\mathcal D_\mathbb C(X)$ the complexification of $\mathcal D(X)$, i.e., the nuclear space of all complex-valued smooth  functions on $X$ with compact support. For $n\ge2$, we denote by $\mathcal D^{\odot n}_\mathbb C(X)$ the $n$th symmetric tensor power of $\mathcal D_\mathbb C(X)$. This is a nuclear space that   consists of all complex-valued symmetric smooth functions on $X^n$ with compact support. We also denote $\mathcal D^{\odot 1}_\mathbb C(X)=\mathcal D_\mathbb C(X)$ and $\mathcal D^{\odot 0}_\mathbb C(X)=\mathbb C$. 

Each function $G^{(n)}\in \mathcal D^{\odot n}_\mathbb C(X)$ can be naturally identified with a function (still denoted by) $G^{(n)}$ on $\Gamma^{(n)}(X)$ by using the formula
$$G^{(n)}(x_1,\dots,x_n)=G^{(n)}(\delta_{x_1}+\dots+\delta_{x_n}),\quad (x_1,\dots,x_n)\in X^{(n)}.$$
Note that, in view of the continuity of the function $G^{(n)}:X^n\to\mathbb C$, the restriction of the function $G^{(n)}$ to $X^{(n)}$ completely determines $G^{(n)}$ on $X^n$. With this identification, we will interpret each $\mathcal D^{\odot n}_\mathbb C(X)$  as the nuclear space  of functions on $\Gamma^{(n)}(X)$. We extend these functions to the whole space $\Gamma_0(X)$ by setting them to be equal to zero on $\Gamma^{(m)}(X)$ for $m\ne n$.

  Let $\Psi$ be the topological direct sum of the nuclear spaces $\mathcal D^{\odot n}_\mathbb C(X)$ with $n\in\mathbb N_0$. Then $\Psi$ is a nuclear space, and it can naturally be interpreted as the space of functions on $\Gamma_0(X)$. Note that $\Psi\subset\mathcal F(\Gamma_0(X))$. In fact, the nuclear space $\Psi$ is densely and continuously embedded into the Hilbert space $\mathcal R_\theta$.  Hence, we get a standard (complex) triple $\Psi\subset \mathcal R_\theta\subset\Psi'$.

The above considerations imply that the real nuclear space $\mathcal D(X)$ is identified with the real nuclear space of functions on $\Gamma^{(1)}(X)$, extended by zero to the whole $\Gamma_0(X)$.  

For each $\varphi\in\mathcal D(X)$, we define a Hermitian operator $A(\varphi)$ in the Hilbert space $\mathcal R_\theta$ with domain $\Psi$ by $A(\varphi)G=\varphi\star G$ for $G\in \Psi$. Note that $A(\varphi)G\in \Psi$.
We now strengthen  condition (A2) by demanding   the following:

\begin{description}

\item[{\rm (A2${}'$)}] For every compact $\Lambda\subset X$, there exists
a constant $C_\Lambda>0$ such that
$\theta(\Gamma_\Lambda^{(n)})\le C_\Lambda^n$ for all $n\in\mathbb N$.
\end{description}

Under the assumptions (A1), (A2${}'$) and (A3), Theorem~\ref{spectral} can be applied to the Hermitian operators $(A(\varphi))_{\varphi\in\mathcal D(X)}$,  the cyclic vector $\Omega=\chi_{0}$, the indicator function of the zero measure, and the standard triples $\mathcal D(X)\subset L^2(X\to\mathbb R,\sigma)\subset \mathcal D'(X)$ and $\Psi\subset \mathcal R_\theta\subset \Psi'$.

However, Theorem~\ref{spectral} yields the spectral measure $\mu$ as a probability measure on $(\mathcal D'(X),\mathcal C(\mathcal D'(X)))$, whereas we are interested in a point process.

Using ideas from \cite{KK}, it was shown in \cite{BKKL} that the spectral measure $\mu$ is, in fact, concentrated on $\Gamma(X)$, provided the following additional assumption is satisfied: 
\begin{description}

\item[{\rm (A4)}] Every compact $\Lambda\subset X$ can be covered
by a finite union of open sets $\Lambda_1,\dots,\Lambda_k$, $k\in\N$,
which have compact closures and satisfy the estimate
$$\theta(\Gamma^{(n)}(\Lambda_i))\le(2+\epsilon)^{-n}\quad\text{for all
$i=1,\dots,k$ and  $n\in\N$},$$
where $\epsilon=\epsilon(\Lambda)>0$.
\end{description}

For the obtained point process $\mu$, the measure $\theta$ is its correlation measure on $\Gamma_0(X)$. Furthermore the obtained unitary operator  $I:\mathcal R_\theta\to L^2(\Gamma(X),\mu)$ is just the restriction of the $K$-transform $K:L^1(\Gamma_0(X),\theta)\to L^1(\Gamma(X),\mu)$ to $\mathcal R_\theta\subset L^1(\Gamma_0(X),\theta)$.   These results imply in particular, the following

\begin{theorem}\label{cydydye6}
Let $X$ be a connected, oriented $C^\infty$ (non-compact) Riemannian manifold. Let $\theta$ be a measure on $(\Gamma_0(X),\mathcal B(\Gamma_0(X)))$ that satisfies assumptions (A1), (A2\,${}'$),  (A3), and (A4). Then there exists a unique point process $\mu$ in $X$ whose correlation measure on $\Gamma_0(X)$ is $\theta$.  
\end{theorem} 

It should be noted that, of the assumptions of Theorem~\ref{cydydye6}, the really difficult one to check is  (A3). We will discuss below  a class of examples where it is indeed possible to check  condition (A3).

\section{Correlation measure of a family of Hermitian operators}\label{jgbiguyifguy}

In this section, we will briefly discuss a result from \cite{LM}, which was a further development  of the studies in \cite{BKKL}. Our presentation of this result follows \cite{AL2} and \cite{AL1}.
 
Let  $X$ be a locally compact Polish space, and let $\sigma$ be a non-atomic Radon measure on $(X,\mathcal B(X))$.  

Let $\mathcal F$ be a separable Hilbert space and let $\mathcal S$ be a dense subspace of $\mathcal F$.  For each $\Delta\in\mathcal B_0(X)$, let  $\rho(\Delta):\mathcal S\to\mathcal S$ be a linear Hermitian operator in $\mathcal F$. We further assume:

\begin{itemize}

\item for any $\Delta_1,\Delta_2\in\mathcal B_0(X)$ with $\Delta_1\cap\Delta_2=\varnothing$, we have $\rho(\Delta_1\cup\Delta_2)=\rho(\Delta_1)+\rho(\Delta_2)$;

\item the operators $\rho(\Delta)$ commute, i.e.,  
$[\rho(\Delta_1),\rho(\Delta_2)]=0$ for any $\Delta_1,\Delta_2\in\mathcal B_0(X)$.
\end{itemize}

Let $\mathcal A$ denote 
 the  (commutative) $*$-algebra generated by $(\rho(\Delta))_{\Delta\in\mathcal B_0(X)}$. Let $\Omega$ be a fixed vector in $\mathcal S$  with $\|\Omega\|_{\mathcal  F}=1$, and let a state $\tau:\mathcal A\to\mathbb C$ be defined by $\tau(a)=(a\Omega,\Omega)_{\mathcal  F}$ for $a\in\mathcal A$. 

In view of formula \eqref{gfxtdsxzetrsresa8}, we define  Wick polynomials in $\mathcal A$ by the following recurrence formula:
\begin{align}
{:}\rho(\Delta){:}&=\rho(\Delta),\notag\\
{:}\rho(\Delta_1)\dotsm \rho(\Delta_{n+1}){:}&=\rho(\Delta_{n+1})\,
{:}\rho(\Delta_1)\dotsm \rho(\Delta_{n}){:}\notag\\
&\quad-\sum_{i=1}^n
{:}\rho(\Delta_1)\dotsm \rho(\Delta_{i-1})\rho(\Delta_i\cap\Delta_{n+1})\rho(\Delta_{i+1})\dotsm \rho(\Delta_n){:}\,,\label{dsaea78}
 \end{align}
 where $\Delta,\Delta_1,\dots,\Delta_{n+1}\in\mathcal B_0(X)$ and $n\in\mathbb N$. 
 It is easy to see that, for each permutation $\pi\in  \mathfrak S_n$,
$${:}\rho(\Delta_1)\cdots \rho(\Delta_n){:} = {:}\rho(\Delta_{\pi(1)})\cdots \rho(\Delta_{\pi(n)}){:}\, .$$

We assume that, for each $n\in\mathbb N$, there exists a symmetric measure $\theta^{(n)}$ on
$X^n$ that is concentrated on $X^{(n)}$  and 
  such that
\begin{equation}\label{6esuw6u61d}
\theta^{(n)}\big(\Delta_1\times\dots\times\Delta_n\big)=\frac1{n!}\,
\tau\big({:}\rho(\Delta_1)\dotsm \rho(\Delta_{n}){:}\big),\quad  \Delta_1,\dots,\Delta_{n}\in\mathcal B_0(X).\end{equation}
Note that, if the measure $\theta^{(n)}$ exists, then it is unique. 
The $\theta^{(n)}$ is called the  $n$th correlation measure of the operators $\rho(\Delta)$. If  $\theta^{(n)}$ has a density $k^{(n)}$ with respect to $\frac1{n!}\sigma^{\otimes n}$, then $k^{(n)}$ is called the  $n$th correlation function of the operators $\rho(\Delta)$.

\begin{theorem}[\!\!\cite{LM}] \label{cdyre6u3}

Let $(\rho(\Delta))_{\Delta\in\mathcal B_0(X)}$ be a family of Hermitian operators in $\mathcal  F$ as above. In particular, these operators have correlation measures $(\theta^{(n)})_{n=1}^\infty$ respective the state $\tau$.  Furthermore, we assume that the following two conditions are satisfied.

{\rm(B1)} For each $\Delta\in\mathcal B_0(X)$, there exists a constant $C_\Delta>0$ such that
$\theta^{(n)}(\Delta^n)\le  C_\Delta^n$ for all $n\in\mathbb N$.

{\rm(B2)} For each $x\in X$ and any sequence
    $\{\Delta_{l}\}_{l\in\mathbb{N}}\subset\mathcal{B}_{0}(X)$ such
    that $\Delta_{l}\downarrow\{x\}$ (i.e.,     $\Delta_1\supset\Delta_2\supset\Delta_3\supset\cdots$ and $\bigcap_{l=1}^\infty\Delta_l=\{x\}$), we have $C_{\Delta_{l}}\rightarrow
    0$ as $l\rightarrow\infty$.

Then the following statements hold.

{\rm (i)}  Let $\mathfrak D=\{a\Omega\mid a\in\mathcal A\}$ and let $\mathfrak F$ denote the closure of $\mathfrak D$ in $\mathcal  F$. Each operator $(\rho(\Delta),\mathfrak D)$ is essentially self-adjoint in $\mathfrak F$, i.e., the closure  of $\rho(\Delta)$, denoted by $\widetilde \rho(\Delta)$,  is a self-adjoint operator in $\mathfrak F$. 

{\rm (ii)} For any $\Delta_1,\Delta_2\in\mathcal B_0(X)$, the operator-valued measures (resolutions of the identity) of the operators $\widetilde \rho(\Delta_1)$ and $\widetilde \rho(\Delta_2)$ commute. 

{\rm (iii)} There exist a unique point process $\mu$ in $X$ and a unique unitary operator\linebreak $U:\mathfrak F\to L^2(\Gamma_X,\mu)$ satisfying $U\Omega=1$ and 
\[
\big(U\rho(\Delta_1)\dotsm \rho(\Delta_{n})\Omega\big)(\gamma)=\gamma(\Delta_1)\dotsm\gamma(\Delta_n)\]
 for any $\Delta_1,\dots,\Delta_{n}\in\mathcal B_0(X)$ ($n\in\mathbb N$). In particular,
\[
\tau\big(\rho(\Delta_1)\dotsm \rho(\Delta_{n})\big)= \int_{\Gamma_X}\gamma(\Delta_1)\dotsm\gamma(\Delta_n)\,\mu(d\gamma).\]

{\rm (iv)} The correlations measures of the point process  $\mu$ are   $(\theta^{(n)})_{n=1}^\infty$. 
  \label{vrsa456ew}
 \end{theorem}

Let us stress that, in Theorem \ref{cdyre6u3}, the correlation measure $\theta$ on $\Gamma_0(X)$ automatically satisfies 
 condition (A3). On the other hand, the most difficult condition of Theorem~\ref{cdyre6u3} is the very existence of the correlation measures $(\theta^{(n)})_{n=1}^\infty$ for a given family of Hermitian operators $(\rho(\Delta))_{\Delta\in\mathcal B_0(X)}$. 

Below we will discuss several examples of application of Theorem~\ref{cdyre6u3}, but first we will recall some definitions related to quasi-free states.

\section{Quasi-free states on the CAR and CCR algebras}\label{vgcgcddt}

In this section, we follow \cite[Section~5.2.3]{BR}, see also \cite{AL2}. 

We start with the case of the CAR algebra. 
Let $\mathcal H$ and $\mathcal  F$ be  separable complex Hilbert spaces. Let $a^+(\varphi)$ and $a^-(\varphi)$ ($\varphi\in\mathcal H$) be bounded linear operators in $\mathcal  F$ such that $a^+(\varphi)$ linearly depends on $\varphi$ and $a^-(\varphi)=(a^+(\varphi))^*$. Let $a^+(\varphi)$ and $a^-(\varphi)$ satisfy the CAR:
\[
\{ a^+(\varphi), a^+(\psi)\}=\{ a^-(\varphi), a^-(\psi)\}=0,\qquad \{a^-(\varphi), a^+(\psi)\}=(\psi,\varphi)_{\mathcal H}.\]
Here $\{A,B\}=AB+BA$ is the anticommutator. Let $\mathbb A$ be the unital $*$-algebra generated by these operators, called the CAR algebra. We define  field operators 
$$b(\varphi)=a^+(\varphi)+a^-(\varphi),\quad \varphi\in\mathcal H.$$
 As easily seen, these operators also generate $\mathbb A$. 

Let  $\tau:\mathbb A\to\mathbb C$ be a state on $\mathbb A$, i.e., $\tau$ is a linear functional on the vector space $\mathbb A$, $\tau(1)=1$, and $\tau(a^*a)\ge0$ for each $£a\in\mathbb A$. The state $\tau$ is completely determined by the functionals $T^{(n)}:\mathcal H^n\to\mathbb C$ $(n\in\mathbb N)$ defined by
\begin{equation}\label{hyfd6swaq3}
T^{(n)}(\varphi_1,\dots,\varphi_n)=\tau\big(b(\varphi_1)\dotsm b(\varphi_n)\big).
\end{equation}
 The state $\tau$ is called quasi-free if 
 \begin{gather}
 T^{(2n-1)}=0,\label{cdst6w5yw}
 \\
T^{(2n)}(\varphi_1,\dots,\varphi_{2n})=\sum (-1)^{\operatorname{Cross}(\nu)}\, T^{(2)}(\varphi_{i_1},\varphi_{j_1})\dotsm T^{(2)}(\varphi_{i_n},\varphi_{j_n}),\quad n\in\mathbb N ,\label{vcfts5y43}\end{gather}
where the summation in \eqref{vcfts5y43} is over all partitions $\nu=\big\{\{i_1,j_1\},\dots,\{i_n,j_n\}\big\}$ of $\{1,\dots,2n\}$ with $i_k<j_k$ ($k=1,\dots,n$) and $\operatorname{Cross}(\nu)$ denotes  the number of all crossings in $\nu$, i.e., the number of all choices of $\{i_k,j_k\},\{i_l,j_l\}\in\nu$ such that $i_k<i_l<j_k<j_l$.

The state $\tau$ is called  gauge-invariant  if, for each $q\in\mathbb C$ with $|q|=1$, we have $T^{(n)}(q\varphi_1,\dots,q\varphi_n)=T^{(n)}(\varphi_1,\dots,\varphi_n)$ for all $\varphi_1,\dots,\varphi_n\in\mathcal H$, $n\in\mathbb N$. 

The state $\tau$ can also be uniquely characterized by the $n$-point functions 
$$S^{(m,n)}:\mathcal H^{m+n}\to\mathbb C,\quad m,n\in\mathbb N_0,\  m+n\ge 1,$$
 defined by
\begin{align}
&S^{(m,n)}(\varphi_m, \dots,\varphi_2,\varphi_1,\psi_1,\psi_2,\dots,\psi_n)\notag\\
&\quad =\tau\big(a^+(\varphi_m)\dotsm a^+(\varphi_2)a^+(\varphi_1)a^-(\psi_1)a^-(\psi_2)\dotsm a^-(\psi_n)\big).
\label{cftesw54} \end{align}
In fact, the  state $\tau$ is gauge-invariant quasi-free if and only if 
\begin{equation}\label{cxtsw64ured}
S^{(m,n)}(\varphi_m\dots,\varphi_1,\psi_1,\dots,\psi_n)=\delta_{m,n}\operatorname{det}\left[S^{(1,1)}(\varphi_i,\psi_j)\right]_{i,j=1,\dots,n}\,. \end{equation}

There exists a one-to-one correspondence between the set of gauge-invariant quasi-free states $\tau$ and the set of all bounded linear operators $K$ in $\mathcal H$ that satisfy $0\le K\le 1$. More exactly, a state $\tau$ on the CAR algebra $\mathbb A$ is gauge-invariant quasi-free if and only if, for some operator $K$ in $\mathcal H$ satisfying  $0\le K\le 1$,  formula \eqref{cxtsw64ured} holds with $S^{(1,1)}(\varphi,\psi)=(K\varphi,\psi)_{\mathcal H}$ for $\varphi,\psi\in\mathcal H$. 

Next, let us briefly discuss the CCR case. As before, let $\mathcal H$ and $\mathcal  F$ be  separable complex Hilbert spaces. Let $\mathcal V$ be a dense subspace of $\mathcal H$, let $D$ be a dense subspace of $\mathcal F$, and let   $a^+(\varphi):D\to D$ and $a^-(\varphi):D\to D$ ($\varphi\in\mathcal V$) be linear operators such that $a^+(\varphi)$ linearly depends on $\varphi\in\mathcal V$ and $a^-(\varphi)=(a^+(\varphi))^*\restriction_D$. Assume $a^+(\varphi)$ and $a^-(\varphi)$ satisfy the CCR:
\[ 
[a^+(\varphi), a^+(\psi)]=[ a^-(\varphi), a^-(\psi)]=0,\qquad [a^-(\varphi), a^+(\psi)]=(\psi,\varphi)_{\mathcal H}.\]
Here $[A,B]=AB-BA$ is the commutator.  Let $\mathbb A$ be the unital $*$-algebra generated by these operators, called the CCR algebra. We define  field operators $b(\varphi)=a^+(\varphi)+a^-(\varphi)$ ($\varphi\in\mathcal V$); these operators also generate $\mathbb A$. 

Let  $\tau:\mathbb A\to\mathbb C$ be a state on $\mathbb A$. The state $\tau$ is completely determined by the functionals $T^{(n)}:\mathcal V^n\to\mathbb C$ $(n\in\mathbb N)$ defined by
\eqref{hyfd6swaq3}. The state $\tau$ is called quasi-free if formula \eqref{cdst6w5yw} holds, and the following counterpart of formula \eqref{vcfts5y43} holds: 
$$T^{(2n)}(\varphi_1,\dots,\varphi_{2n})=\sum  T^{(2)}(\varphi_{i_1},\varphi_{j_1})\dotsm T^{(2)}(\varphi_{i_n},\varphi_{j_n}),\quad n\in\mathbb N .$$

\begin{remark}\label{vtydfytdr6e6} Let $\varkappa:\mathcal V\to\mathbb C$ be a linear functional. Assume that the operators $a^+(\varphi)$, $a^-(\varphi)$ ($\varphi\in\mathcal V$) satisfy the CCR. Then the operators $A^+(\varphi)=a^+(\varphi)+\varkappa(\varphi)$, $A^-(\varphi)=a^-(\varphi)+\overline{\varkappa(\varphi)}$ ($\varphi\in\mathcal V$) also satisfy the CCR. Note that the unital $*$-algebra $\mathbb A$ generated by the operators  $a^+(\varphi)$, $a^-(\varphi)$ coincides with the unital $*$-algebra generated by the operators  $A^+(\varphi)$, $A^-(\varphi)$. Assume that $\tau$ is a quasi-free state on $\mathbb A$ relative the operators $a^+(\varphi)$, $a^-(\varphi)$. But then $\tau$ is not a quasi-free state on $\mathbb A$ relative the operators $A^+(\varphi)$, $A^-(\varphi)$. Indeed, in the latter case, $T^{(2n-1)}\ne0$. Nevertheless, one can easily generalize the definition of a quasi-free state  in such a way that, if $\tau$ is a quasi-free state relative the operators $a^+(\varphi)$, $a^-(\varphi)$, then 
$\tau$ is also a quasi-free state relative the operators $A^+(\varphi)$, $A^-(\varphi)$.
\end{remark}

Similarly to the CAR case, we define a gauge-invariant state $\tau$ on $\mathbb A$ and $n$-point functions 
$$S^{(m,n)}:\mathcal V^{m+n}\to\mathbb C,\quad m,n\in\mathbb N_0,\  m+n\ge 1,$$
by formula \eqref{cftesw54}.

A state $\tau$ is gauge-invariant quasi-free if and only if the formula \eqref{cxtsw64ured} holds in which the determinant $\operatorname{det}$ is replaced with the permanent $\operatorname{per}$.  In fact, a state $\tau$ on the CCR algebra $\mathbb A$ is gauge-invariant quasi-free if and only if, for a bounded linear operator $K$ in $\mathcal H$ satisfying  $K\ge0$,  we have
 $$
 S^{(m,n)}(\varphi_m\dots,\varphi_1,\psi_1,\dots,\psi_n)=\delta_{m,n}\operatorname{per}\left[(K\varphi_i,\psi_j)_{\mathcal H}\right]_{i,j=1,\dots,n}\,. $$

Following \cite{AL2} (see also the references therein), let us now present a formal observation. 
Choose $\mathcal H=L^2(X,\sigma)$, and in the CCR case, assume that $\mathcal V$ contains all bounded functions with compact support. Let $a^+(\varphi)$ and $a^-(\varphi)$  satisfy either the CAR or the CCR, and let $\tau$ be a state on the corresponding $*$-algebra $\mathbb A$. For $x\in X$, define (formal) operators $a^+(x)$, $a^-(x)$ that satisfy
$$a^+(\chi_\Delta)=\int_\Delta a^+(x)\,\sigma(dx),\quad a^-(\chi_\Delta)=\int_\Delta a^-(x)\,\sigma(dx)\quad\text{for each }\Delta\in\mathcal B_0(X). $$
(Here $\chi_\Delta$ denotes the indicator of $\Delta$.) We now define particle density $\rho(x)=a^+(x)a^-(x)$ ($x\in X$), and in the smeared form
$$\rho(\Delta)=\int_\Delta\rho(x)\,\sigma(dx)=\int_\Delta a^+(x)a^-(x)\,\sigma(dx),\quad \Delta\in\mathcal B_0(X).$$
It follows from the CAR/CAR that the Hermitian operators $\rho(\Delta)$ ($\Delta\in\mathcal B_0(X)$) commute, and furthermore, formula \eqref{dsaea78} implies, for any $\Delta_1,\dots,\Delta_{n}\in\mathcal B_0(X)$,
\begin{equation}\label{terw5yw35}
{:}\rho(\Delta_1)\dotsm \rho(\Delta_{n}){:}=\int_{\Delta_1\times\dots\times \Delta_n}a^+(x_n)\dotsm a^+(x_1)a^-(x_1)\dotsm a^-(x_n)\,\sigma^{\otimes n}(dx_1\dotsm dx_n).\end{equation}
Thus, the Wick polynomials correspond to the Wick (normal) ordering, in which all the  operators $a^+(x_i)$ are to the left of  all the operators $a^-(x_j)$. 
Hence, by \eqref{6esuw6u61d} and~\eqref{terw5yw35}, we formally obtain
\begin{align*}
&\theta^{(n)}\big(\Delta_1\times\dots\times\Delta_n\big)\notag\\
&\quad =\frac1{n!}\int_{\Delta_1\times\dots\times \Delta_n}\tau\big(a^+(x_n)\dotsm a^+(x_1)a^-(x_1)\dotsm a^-(x_n)\big)\sigma^{\otimes n}(dx_1\dotsm dx_n).\end{align*}
Therefore, by \eqref{5w738}, the family of the operators $(\rho(\Delta))_{\Delta\in\mathcal B_0(X)}$ has correlation functions 
$$k^{(n)}(x_1,\dots,x_n)=\tau\big(a^+(x_n)\dotsm a^+(x_1)a^-(x_1)\dotsm a^-(x_n)\big).$$

Below we will present several examples where a rigorous representation of the CAR/CCR leads, through an informal particle density $\rho(x)$ ($x\in X$), to a rigorously defined family of Hermitian operators $(\rho(\Delta))_{\Delta\in\mathcal B_0(X)}$ that satisfy the conditions of Theorem~\ref{cdyre6u3}, and hence a point process $\mu$ exists and is the spectral measure of the family of commuting self-adjoint operators $(\tilde\rho(\Delta))_{\Delta\in\mathcal B_0(X)}$.   But first we will recall  several basic facts about antisymmetric and symmetric Fock spaces.

\section{Fock spaces} \label{fxdsdese}

The material of this section is based on \cite[Chapter~2, Section~2.2]{BK} and \cite[Section~19]{Parth}

Let $\mathcal G$ denote a separable complex Hilbert space. Let 
$\mathcal{AF}(\mathcal G)=\bigoplus_{n=0}^\infty \mathcal G^{\wedge n}n!$ 
denote the  antisymmetric Fock space over $\mathcal G$. Here $\wedge$ denotes the antisymmetric tensor product and elements of the Hilbert space $\mathcal{AF}(\mathcal G)$ are sequences $g=(g^{(n)})_{n=0}^\infty$ with $g^{(n)}\in  \mathcal G^{\wedge n}$ ($ \mathcal G^{\wedge 0}=\mathbb C$) and $\|g\|^2_{\mathcal{AF}(\mathcal G)}=\sum_{n=0}^\infty \|g^{(n)}\|^2_{\mathcal G^{\wedge n}}\,n!<\infty$. The vector $\Omega=(1,0,0,\dots)$ is called the  vacuum.

For $\varphi\in\mathcal G$, we define a creation operator $a^+(\varphi)$ as a bounded linear operator in $\mathcal{AF}(\mathcal G)$ that satisfies $a^+(\varphi)g^{(n)}=\varphi\wedge g^{(n)}$ for $g^{(n)}\in\mathcal G^{\wedge n}$. 
 For each $\varphi\in\mathcal G$, we define an  annihilation operator  $a^-(\varphi)=a^+(\varphi)^*$. Then, 
 $$a^-(\varphi)g_1\wedge\dotsm \wedge g_n=\sum_{i=1}^n (-1)^{i+1}(g_i,\varphi)_{\mathcal G}\,g_1\wedge\dotsm \wedge g_{i-1}\wedge g_{i+1}\wedge \dotsm \wedge g_n$$
 for all $g_1,\dots,g_n\in\mathcal G$.
 Note that the norm of the  operators $a^+(\varphi)$, $a^-(\varphi)$  in $\mathcal{AF}(\mathcal G)$  is equal to $\|\varphi\|_{\mathcal G}$.
The operators $a^+(\varphi)$, $a^-(\varphi)$ satisfy the CAR over $\mathcal G$.

We will also need below the differential second quantization. We denote by $\mathcal{AF}_{\mathrm{fin}}(\mathcal G)$ the subspace of $\mathcal{AF}(\mathcal G)$ that consists of all $(g^{(n)})_{n=0}^\infty \in \mathcal{AF}(\mathcal G)$ such that, for some $N\in\mathbb N$, $g^{(n)}=0$ for all $n>N$. For a bounded linear operator $A$ in $\mathcal G$, its differential second quantization $d\Gamma(A)$ is the linear operator in $\mathcal{AF}(\mathcal G)$, with domain $\mathcal{AF}_{\mathrm{sym}}(\mathcal G)$, that maps each subspace $\mathcal G^{\wedge n}$ continuously into itself and satisfies $d\Gamma(A)\Omega=0$ and 
$$d\Gamma(A)g_1\wedge\dotsm \wedge g_n=\sum_{i=1}^n g_1\wedge\dotsm \wedge g_{i-1}\wedge(Ag_i)\wedge g_{i+1}\dotsm \wedge g_n$$
 for all $g_1,\dots,g_n\in\mathcal G$.
 
 Next, let 
$\mathcal{SF}(\mathcal G)=\bigoplus_{n=0}^\infty \mathcal G^{\odot n}n!$ 
denote the  symmetric Fock space of $\mathcal G$. Here $\odot$ denotes the symmetric tensor product and elements of the Hilbert space $\mathcal{SF}(\mathcal G)$ are sequences $g=(g^{(n)})_{n=0}^\infty$ with $g^{(n)}\in  \mathcal G^{\odot n}$ ($ \mathcal G^{\odot 0}=\mathbb C$) and $\|g\|^2_{\mathcal{SF}(\mathcal G)}=\sum_{n=0}^\infty \|g^{(n)}\|^2_{\mathcal G^{\odot n}}\,n!<\infty$. The vector $\Omega=(1,0,0,\dots)$ is again called the  vacuum. Let also $\mathcal{SF}_{\mathrm{fin}}(\mathcal G)$ be the subspace of $\mathcal{SF}(\mathcal G)$ defined similarly to  $\mathcal{AF}_{\mathrm{fin}}(\mathcal G)$.

For $\varphi\in\mathcal G$, we define a creation operator $a^+(\varphi)$ as a linear operator in $\mathcal{SF}(\mathcal G)$ with domain $\mathcal{SF}_{\mathrm{fin}}(\mathcal G)$ 
that satisfies $a^+(\varphi)g^{(n)}=\varphi\odot g^{(n)}$ for $g^{(n)}\in\mathcal G^{\odot n}$. 
 For each $\varphi\in\mathcal G$, we define an  annihilation operator  $a^-(\varphi)$ as the adjoint of the operator $a^+(\varphi)$ restricted to $\mathcal{SF}_{\mathrm{fin}}(\mathcal G)$. Then, 
 $$a^-(\varphi)g_1\odot\dotsm \odot g_n=\sum_{i=1}^n (g_i,\varphi)_{\mathcal G}\,g_1\odot\dotsm \odot g_{i-1}\odot g_{i+1}\odot \dotsm \odot g_n$$
 for all $g_1,\dots,g_n\in\mathcal G$.
 The operators $a^+(\varphi)$, $a^-(\varphi)$ satisfy the CAR over $\mathcal G$.

Next, we define a differential second quantization in $\mathcal{SF}(\mathcal G)$.
 For a bounded linear operator $A$ in $\mathcal G$, its differential second quantization $d\Gamma(A)$ is a linear operator in $\mathcal{SF}(\mathcal G)$, with domain $\mathcal{SF}_{\mathrm{sym}}(\mathcal G)$, that  maps each subspace $\mathcal G^{\odot  n}$ continuously  into itself and satisfies   $d\Gamma(A)\Omega=0$ and 
$$d\Gamma(A)g_1\odot\dotsm \odot g_n=\sum_{i=1}^n g_1\odot\dotsm \odot g_{i-1}\odot(Ag_i)\odot g_{i+1}\odot\dotsm \odot g_n$$
 for all $g_1,\dots,g_n\in\mathcal G$.

\section{Examples of application of Theorem~\ref{cdyre6u3}}\label{bfydr7er74r}
Everywhere below $X$ is a locally compact Polish space and $\sigma$ is a non-atomic Radon measure on $X$.  
\subsection{Determinantal point process with a Hermitian kernel}\label{vgfhxdtdstd}

We discuss here a result from \cite{LM}, see also \cite{L2}.

First, let us briefly recall Araki--Wyss' explicit construction of the gauge-invariant quasi-free states on the CAR algebra  \cite{AW}. 
Let  $\mathcal H$ be a separable complex Hilbert space with an antiunitary involution $\mathcal I$, i.e., $\mathcal I$ is an antilinear operator in $\mathcal H$ that satisfies $(\mathcal I\varphi,\mathcal I\psi)_{\mathcal H}=(\psi,\varphi)_\mathcal H$ for $\varphi,\psi\in\mathcal H$ and $\mathcal I^2=\mathbf 1$. 
Let $\mathcal G=\mathcal H\oplus \mathcal H$, and for $\varphi\in\mathcal H$ and $\sharp\in\{+,-\}$, we define operators $a^\sharp _1(\varphi)=a^\sharp (\varphi,0)$, $a^\sharp _2(\varphi)=a^\sharp(0,\varphi)$ in $\mathcal{AS}(\mathcal G)$. 

We fix a bounded linear operator $K$ in $\mathcal H$ that satisfies  $ 0\le K\le 1$, and define operators $K_1=\sqrt K$ and $K_2=\sqrt{1-K}$.  For each $\varphi\in\mathcal H$, we define the following bounded linear operators in $\mathcal{AF}(\mathcal G)$:
\begin{equation}\label{css}
 A^+(\varphi)=a^+_2( K_2\varphi)+a_1^-(\mathcal I K_1\varphi),\quad 
A^-(\varphi)=a_2^-( K_2\varphi)+a_1^+(\mathcal I K_1\varphi).
\end{equation}
 The operators $ A^+(\varphi)$, $ A^-(\varphi)$ ($\varphi\in\mathcal H$) satisfy the CAR. 
 
  Let $\mathbb A$ denote the corresponding CAR $*$-algebra. The vacuum state on $\mathbb A$ is defined by $\tau(a)=(a\Omega,\Omega)_{\mathcal {AF}(\mathcal G)}$ ($a\in\mathbb A$). This state is gauge-invariant quasi-free,  with 
  \begin{equation}\label{vgniuguyfucydy}
  S^{(1,1)}(\varphi,\psi)=(K\varphi,\psi)_{\mathcal H},\quad \varphi,\psi\in\mathcal H.
  \end{equation}
  
  Now, let $\mathcal H=L^2(X,\sigma)$ with $\mathcal I$ being the complex conjugation in $\mathcal H$. For each $\Delta\in\mathcal B_0(X)$, we denote by $P_\Delta$ the orthogonal projection of $L^2(X,\sigma)$ onto its subspace $L^2(\Delta,\sigma)$, i.e., for $f\in L^2(X,\sigma)$, $(P_\Delta f)(x)=\chi_\Delta(x)f(x)$, where $\chi_\Delta$ is the indicator function of the set $\Delta$. 
  
 Let $K$ be a bounded linear operator in $\mathcal H$ that satisfies  $ 0\le K\le 1$, and additionally we assume that the operator $K$ is locally trace-class, i.e., for each $\Delta\in\mathcal B_0(X)$, the operator $P_\Delta KP_\Delta$ is trace-class.  Under these assumptions,  $K$ is an integral  operator, and furthermore its integral kernel $K(x,y)$ can be chosen so that $\operatorname{Tr}(P_\Delta KP_\Delta)=\int_\Delta K(x,x)\,\sigma(dx)$ for each $\Delta\in\mathcal B_0(X)$.   
 
 Under these assumptions, the particle density corresponding to the gauge-invariant quasi-free  
(vacuum) state $\tau$ on $\mathbb A$ is the family of (unbounded) Hermitian operators $(\rho(\Delta))_{\Delta\in\mathcal B_0(X)}$ that are defined as follows. 

Let $\mathcal H_1$ and $\mathcal H_2$ denote two copies of $\mathcal H$, so that 
$$\mathcal G^{\otimes 2}=(\mathcal H_1\oplus\mathcal H_2)^{\otimes2}=\bigoplus_{i,j\in\{1,2\}}\mathcal H_i\otimes\mathcal H_j.$$
 Let $\mathcal E_{i,j}:\mathcal H^{\otimes 2}\to\mathcal G^{\otimes 2}$ denote the operator of embedding of $\mathcal H^{\otimes 2}$ into $\mathcal G^{\otimes 2}$ under which and element $f^{(2)}\in\mathcal H^{\otimes 2}$ is mapped to $f^{(2)}\in\mathcal H_i\otimes\mathcal H_j\subset\mathcal G^{\otimes 2}$.  We denote $(f^{(2)})_{i,j}=\mathcal E_{i,j}f^{(2)}$. 
 
 Next, for $g^{(2)}\in\mathcal G^{\otimes 2}$ we define a creation operator $a^+(g^{(2)})$ which satisfies
 \begin{equation}\label{gcfdxtsteste5}
 a^+(g^{(2)})\varphi^{(n)}=\operatorname{ASym}_{n+2}(g^{(2)}\otimes \varphi^{(n)}),\quad \varphi^{(n)}\in \mathcal G^{\wedge n},\ n\in\mathbb N_0 ,
 \end{equation}
 where $\operatorname{ASym}_{n+2}$ is the operator of antisymmetrization that projects $\mathcal G^{\otimes (n+2)}$ onto $\mathcal G^{\wedge(n+2)}$.  The operator $a^+(g^{(2)})$ is bounded in $\mathcal{AF}(\mathcal G)$ and we denote by $a^-(g^{(2)})$ its adjoint operator. In particular, for each $f^{(2)}\in\mathcal H^{\otimes 2}$ and $i,j\in\{1,2\}$, we have defined bounded linear operators $a^+((f^{(2)})_{i,j})$ and $a^-((f^{(2)})_{i,j})$ in $\mathcal{AF}(\mathcal G)$.

For each $\Delta\in\mathcal B_0(X)$,  $K_2P_\Delta K_1$ is a Hilbert--Schmidt operator in  $\mathcal H$, hence it can be standardly identified with an element of $\mathcal H^{\otimes 2}$.     Then, for each $\Delta\in\mathcal B_0(X)$, the particle density $\rho(\Delta)$ is the linear Hermitian operator in $\mathcal{AF}(\mathcal G)$ with domain $\mathcal{AF}_{\mathrm{fin}}(\mathcal G)$ given by
\begin{align}
\rho(\Delta)&=a^+((K_2P_\Delta K_1)_{2,1})+a^-((K_2P_\Delta K_1)_{2,1})\notag\\
&\quad+d\Gamma\big((-\mathcal IK_1P_\Delta K_1\mathcal I)\oplus (K_2 P_\Delta K_2)\big)+\operatorname{Tr}(P_\Delta KP_\Delta).
\label{vtdrdsr} 
\end{align}

The family of the Hermitian operators $(\rho(\Delta))_{\Delta\in\mathcal B_0(X)}$ satisfies the conditions of Theorem~\ref{cdyre6u3}. The corresponding point process $\mu$ has correlation functions
$$k^{(n)}(x_1,\dots,x_n)=\operatorname{det}[K(x_i,x_j)]_{i,j=1,\dots,n}\,.$$
Thus, $\mu$ is  the determinantal point process whose (Hermitian) correlation kernel is $ K(x,y)$.   

\subsection{Determinantal point process with a $J$-Hermitian kernel}

We discuss here a result from \cite{AL1}, see also the references therein.

Assume that the underlying space $X$ is split into two disjoint measurable parts, $X_1$ and $X_2$, of positive measure $\sigma$. We denote by $P_i$  the orthogonal projection of $\mathcal H=L^2(X,\sigma)$ onto $\mathcal H_i=L^2(X_i,\sigma)$, and we define  $J=P_1-P_2$.

  According to the orthogonal sum $\mathcal H=\mathcal H_1\oplus \mathcal H_2$, each bounded linear operator $A$ in $\mathcal H$ can be represented 
  in the block form, \[
A=\left[\begin{matrix}
A^{11}& A^{21}\\
A^{12}&A^{22}
\end{matrix}\right],\]
where $A^{ij}=P_iAP_j$.  

One says that a bounded linear  operator $A$ in $\mathcal H$ is $J$-self-adjoint if  $(A^{ii})^*=A^{ii}$ ($i=1,2$) and $(A^{21})^*=-A^{12}$. If a $J$-self-adjoint operator $A$ is an integral operator, then its integral kernel $A(x,y) $ is called $J$-Hermitian. A $J$-Hermitian kernel $A(x,y)$ satisfies $A(x,y)=\overline{A(y,x)}$ if both $x$ and $y$ are from the same part $X_i$, and $A(x,y)=-\overline{A(y,x)}$ if  $x$ and $y$ are from  different parts $X_i$ and $X_j$ ($i\ne j$). 

For a bounded linear operator $A$ in $\mathcal H$, we denote $\widehat A=AP_1+(1-A)P_2$. As easily seen, the operation $A\mapsto\widehat A$ is an involution in the space of bounded linear operators in $\mathcal H$. Furthermore, if an operator $A$ is self-adjoint, then $\widehat A$ is $J$-self-adjoint, and if $A$ is $J$-self-adjoint, then $\widehat A$ is self-adjoint. 

Below we will use some of the notations of Section \ref{vgfhxdtdstd}. Let  $K$ be a bounded linear operator in $\mathcal H$ with $0\le K\le 1$, and let $K_1=\sqrt{K}$, $K_2=\sqrt{1-K}$.  For each $\varphi\in\mathcal H$, consider the following bounded linear operators in $\mathcal{AF}(\mathcal G)$:
\begin{align*}
A^+(\varphi)&=a^+(\mathcal IK_1\mathcal IP_2 \varphi,\, K_2P_1\varphi)+a^-(\mathcal I K_1P_1\varphi,\,\mathcal IK_2 P_2 \varphi),\\
A^-(\varphi)&=a^-(\mathcal IK_1\mathcal IP_2\varphi,\,K_2P_1\varphi)+a^+(\mathcal IK_1P_1\varphi,\mathcal I K_2P_2\varphi).
\end{align*}
These operators  satisfy the CAR, and  let $\mathbb A$ be the corresponding CAR $*$-algebra. 
The vacuum state $\tau$ on $\mathbb A$ is not gauge-invariant but it is still quasi-free.
In fact, formulas \eqref{cdst6w5yw}, \eqref{vcfts5y43} hold in this case with 
$$T^{(2)}(\varphi,\psi)=2i\Im( K\mathbb J\varphi,\mathbb J\psi)_{\mathcal H}+(\mathbb J\psi,\mathbb J\varphi)_{\mathcal H},\quad \varphi,\psi\in\mathcal H,$$
where $\mathbb J\varphi=P_1\varphi+P_2\mathcal I\varphi$.

Let us now  additionally assume that both operators $P_1KP_1$ and $P_2(1-K)P_2$ are locally trace-class. Then the corresponding particle density can be rigorously defined as follows. For each $\Delta\in\mathcal B_0(X)$, the particle density $\rho(\Delta)$ is the linear Hermitian operator in $\mathcal{AF}(\mathcal G)$ with domain $\mathcal{AF}_{\mathrm{fin}}(\mathcal G)$ given by
\begin{align*}
\rho(\Delta)=&a^+\big(\big(K_2J_{\Delta} K_1\big)_{2,1}\big)+a^-\big(\big(K_2J_{\Delta} K_1\big)_{2,1}\big)+d\Gamma \big((-\mathcal I K_1J_{\Delta} K_1\mathcal I)\oplus (K_2J_{\Delta} K_2)\big)\\
&+ \operatorname{Tr}(P_{\Delta\cap X_1}KP_{\Delta\cap X_1})+ \operatorname{Tr}(P_{\Delta\cap X_2}(1-K)P_{\Delta\cap X_2}),
\end{align*}
where  $J_\Delta=P_{\Delta\cap X_1}-P_{\Delta\cap X_2}$. Here $(K_2J_{\Delta} K_1)_{2,1}\in\mathcal G^{\otimes 2}$ is defined similarly to ($K_2P_\Delta K_1)_{2,1}$ in 
Section \ref{vgfhxdtdstd}. 

The family of the Hermitian operators $(\rho(\Delta))_{\Delta\in\mathcal B_0(X)}$ satisfies the conditions of Theorem~\ref{cdyre6u3}. The corresponding point process $\mu$ has correlation functions
$$k^{(n)}(x_1,\dots,x_n)=\operatorname{det}[\mathbb K(x_i,x_j)]_{i,j=1,\dots,n}.$$
Here $\mathbb K=\widehat K$ is a $J$-self-adjoint operator in $\mathcal H$, and $\mathbb K(x,y)$ is its integral $J$-Hermitian kernel. Note that $\mathbb K(x,y)$ indeed exists due to our assumptions, and furthermore $\mathbb K(x,y)$ is chosen so that $\operatorname{Tr}(P_\Delta KP_\Delta)=\int_\Delta\mathbb K(x,x)\,\sigma(dx)$ for each $\Delta\in\mathcal B_0(X)$ with $\Delta\subset X_1$ and $\operatorname{Tr}(P_\Delta (1-K)P_\Delta)=\int_\Delta\mathbb K(x,x)\,\sigma(dx)$ for each $\Delta\in\mathcal B_0(X)$ with $\Delta\subset X_2$.

 Thus, $\mu$ is a determinantal point process with a $J$-Hermitian correlation kernel $\mathbb K(x,y)$. 

\subsection{Poisson point process}

We will discuss here a result from \cite{AL2}, compare with \cite{Ber}.

Let $\mathcal H=L^2(X,\sigma)$ and let $\mathcal V=L^2(X,\sigma)\cap L^1(X,\sigma)$. For each $\varphi\in\mathcal V$, we define the following linear operators in $\mathcal{SF}(\mathcal H)$ with domain $\mathcal{SF}_{\mathrm{fin}}(\mathcal H)$:
$$A^+(\varphi)=a^+(\varphi)+\int_X \varphi(x)\,\sigma(dx),\quad A^-(\varphi)=a^-(\varphi)+\int_X \varphi(x)\,\sigma(dx).$$
These operators satisfy the CCR. The vacuum state on the corresponding CCR algebra $\mathbb A$ is quasi-free in the sense as explained in Remark~\ref{vtydfytdr6e6}.

The corresponding particle density has the form
$$\rho(\Delta)=a^+(\chi_\Delta)+a^-(\chi_\Delta)+d\Gamma(P_\Delta)+\sigma(\Delta),$$
where $P_\Delta$ is the orthogonal projection of $\mathcal H$ onto  $L^2(\Delta,\sigma)$.

The family of the Hermitian operators $(\rho(\Delta))_{\Delta\in\mathcal B_0(X)}$ satisfies the conditions of Theorem~\ref{cdyre6u3}. The corresponding point process $\mu$ has correlation functions $k^{(n)}(x_1,\dots,x_n)=1$ for all $n\in\mathbb N$. Thus, $\mu$ is the Poisson point process with intensity measure $\sigma$. 

\subsection{Permanental point process}
We discuss here a result from \cite{LM}, see also \cite{L2}.

The results of this section are pretty similar to those of Section \ref{vgfhxdtdstd}, so we will only outline the necessary changes. 

First, let us briefly recall Araki--Woods' explicit construction of the gauge-invariant quasi-free states on the CCR algebra  \cite{ArWoods}. 
 
 We fix a bounded linear operator $K$ in $\mathcal H$ that satisfies  $ K\ge0$, and define operators $K_1=\sqrt K$ and $K_2=\sqrt{1+K}$.  For each $\varphi\in\mathcal H$, we define the linear operators in $\mathcal{SF}(\mathcal G)$ with the domain $\mathcal{SF}_{\mathrm{fin}}(\mathcal G)$ through formula \eqref{css}, with the corresponding operators acting in $\mathcal{SF}(\mathcal G)$  rather than $\mathcal{AF}(\mathcal G)$. These operators satisfy the CCR, and the vacuum state on the corresponding CCR algebra is gauge-invariant quasi-free, and \eqref{vgniuguyfucydy} holds. 
 
Let now $\mathcal H=L^2(X,\sigma)$ and assume that the operator $K\ge 0$ is locally trace-class. For $g^{(2)}\in\mathcal G^{\otimes 2}$, we define  operators $a^+(g^{(2)})$, $a^-(g^{(2)})$ similarly to Section \ref{vgfhxdtdstd}. In particular, to define $a^+(g^{(2)})$, we use formula \eqref{gcfdxtsteste5} in which we replace $\mathcal G^{\wedge n}$ with $\mathcal G^{\odot n}$, and the antysimmetrization operator $\operatorname{ASym}_{n+2}$ with the symmetrization operator $\operatorname{Sym}_{n+2}$.  Next, 
  we define the corresponding particle density $\rho(\Delta)$ similarly to  formula \eqref{vtdrdsr}. The family of the Hermitian operators $(\rho(\Delta))_{\Delta\in\mathcal B_0(X)}$ satisfies the conditions of Theorem~\ref{cdyre6u3}. The corresponding point process $\mu$ has correlation functions
$$k^{(n)}(x_1,\dots,x_n)=\operatorname{per}[K(x_i,x_j)]_{i,j=1,\dots,n}.$$
Thus, $\mu$ is  the permanental  point process whose correlation kernel is $ K(x,y)$.   

 \subsection{Hafnian point process}
 
 In this section we will briefly discuss a result from \cite{AL2},  see also the references therein.
 
 Let, as before, $\mathcal H=L^2(X,\sigma)$, and let $\mathcal V$ be the (dense) subspace of $\mathcal H$ that consists of all bounded measurable functions with compact support.  Let $\mathcal G$ be a separable complex Hilbert space with an antiunitary involution $\mathcal I$. 
   
 Let $X\ni x\mapsto (L_1(x),L_2(x))\in\mathcal G^2$ be a measurable mapping. We assume that
\begin{align}
&(L_1(x),\mathcal IL_2(y))_{\mathcal G}=(L_1(y),\mathcal IL_2(x))_{\mathcal G},\notag\\
&(L_1(x),L_1(y))_{\mathcal G}=(L_2(x),L_2(y))_{\mathcal G}\quad \text{for all }x,y\in X,\notag\\
&\int_\Delta\|L_1(x)\|_{\mathcal G}^2\,\sigma(dx)=\int_\Delta\|L_2(x)\|_{\mathcal G}^2\,\sigma(dx)<\infty\quad \text{for each }\Delta\in\mathcal B_0(X). \label{xzra4yrde4}
\end{align}
  By using e.g.\ \cite[Chapter 10, Theorem~3.1]{BUS},
 for each $h\in\mathcal V$, we define
\begin{equation*}
\int_X h L_i\,d\sigma,\ \int_X h\,\mathcal IL_i\,d\sigma\in\mathcal G,\quad i=1,2,\end{equation*}
as  Bochner integrals. 

Denote $\mathcal E=\mathcal H\oplus \mathcal G$. We consider   the following linear operators in $\mathcal {SF}(\mathcal E)$ with domain $\mathcal {SF}_{\mathrm{fin}}(\mathcal E)$:
\begin{align}
A^+(h)=&a^+\bigg(h,\int_X h\,\mathcal IL_2\,d\sigma\bigg)+a^-\bigg(0,\int_X \overline h\,L_1\,d\sigma\bigg),\notag\\
A^-(h)=&a^-\bigg(h,\int_X h\,\mathcal IL_2\,d\sigma\bigg)+a^+\bigg(0,\int_X \overline h\,L_1\,d\sigma\bigg),\quad h\in\mathcal V.\label{tdr65eS}
\end{align}
These operators satisfy the CCR, and the vacuum state $\tau$ on the corresponding CCR algebra $\mathbb A$ is quasi-free with
\begin{align}
T^{(2)}(f,h)&=\int_X \overline{f(x)}\, h(x)\sigma(dx)\notag\\
&\quad+2\int_{X^2}\Re\left(f(x)h(y)\overline{\mathcal K_2(x,y)}+\overline{f(x)}\,h(y)\mathcal K_1(x,y)\right)\sigma^{\otimes 2}(dx\,dy).\notag
\end{align} 
Here
\begin{align*}
\mathcal K_1(x,y)&=(L_1(x),L_1(y))_{\mathcal G},\\ 
\mathcal K_2(x,y)&=(L_1(x),\mathcal IL_2(y))_{\mathcal G},\quad x,y\in X.
\end{align*}
The vacuum state $\tau$ is, however, not gauge-invariant. 

It follows from \eqref{tdr65eS} that the operators $A^+(x)$, $A^-(x)$ ($x\in X$) are formally given by 
\begin{align}
A^+(x)&=a^+(x,0)+a^+(0,\mathcal IL_2(x))+a^-(0,L_1(x)),\notag\\
A^-(x)&=a^-(x,0)+a^-(0,\mathcal IL_2(x))+a^+(0,L_1(x)). \label{ytyuryr}\end{align}
Note that, in  formula \eqref{ytyuryr}, $a^\sharp(0,\mathcal IL_2(x))$ and $a^\sharp(0,L_1(x))$ ($\sharp\in\{+,-\}$) are rigorously defined operators in $\mathcal{SF}(\mathcal E)$ with domain $\mathcal {SF}_{\mathrm{fin}}(\mathcal E)$. On the other hand, $a^\sharp(x,0)$ are formal operators that satisfy
$$\int_\Delta  a^\sharp(x,0)\,\sigma(dx)=a^\sharp (\chi_\Delta,0),\quad \Delta\in\mathcal B_0(X),\ \sharp\in\{+,-\}. $$
 In \cite[Proposition~6.4]{AL2}, it is shown that the particle density $\rho(\Delta)=\int_\Delta A^+(x)A^-(x)\,\sigma(dx)$ ($\Delta\in\mathcal B_0(X)$) can be easily given a rigorous meaning as a Hermitian operator in $\mathcal{SF}(\mathcal E)$ with domain $\mathcal {SF}_{\mathrm{fin}}(\mathcal E)$. 
 
The family  $(\rho(\Delta))_{\Delta\in\mathcal B_0(X)}$ satisfies the conditions of Theorem~\ref{cdyre6u3}. 
To write down the correlation functions of the corresponding point process $\mu$, let us first recall that, for  
a symmetric $2n\times2n$-matrix $C=[c_{ij}]_{i,j=1,\dots,2n}$, the  hafnian of $C$ is defined by
$$\operatorname{haf}(C)=\frac1{n!\,2^n}\sum_{\pi\in\mathfrak S_{2n}}\prod_{i=1}^n c_{\pi(2i-1)\pi(i)}. $$
(Note the the value of the hafnian of $C$ does not depend on the diagonal elements  of the matrix $C$.)
The hafnian can also be written as
\begin{equation*}
\operatorname{haf}(C)=\sum c_{i_1j_1}\dotsm c_{i_nj_n},\end{equation*}
where the summation is over all partitions $\nu=\big\{\{i_1,j_1\},\dots,\{i_n,j_n\}\big\}$ of $\{1,\dots,2n\}$. 
The point process $\mu$ has correlation functions  
 $$k^{(n)}(x_1,\dots,x_n)=\operatorname{haf}\big[\mathbb K(x_i,x_j)\big]_{i,j=1,\dots,n}\,,$$
 where 
\begin{equation*}
\mathbb K(x,y)=\left[\begin{matrix}
\mathcal K_2(x,y)&\mathcal K_1(x,y)\\\overline{\mathcal K_1(x,y)}&\overline{\mathcal K_2(x,y)}\end{matrix}\right].\end{equation*}
Hence, it is natural to call $\mu$ a hafnian point process. 

In fact, $\mu$ is a Cox process. More exactly, $\mu$ is the Poisson point process with random intensity measure $|G(x)|^2\sigma(dx)$. Here $(G(x))_{x\in X}$ is a complex-valued Gaussian random field on $X$ with covariance  $\mathbb E(G(x)\overline{G(y)})=\mathcal K_1(x,y)$ and pseudo-covariance $\mathbb E(G(x)G(y))=\mathcal K_2(x,y)$.

Let us now consider a special case of the above construction. Let $K\ge0$ be a  locally trace-class operator in $\mathcal H$. Let $K_1=\sqrt{K}$. Then, for each $\Delta\in\mathcal B_0(X)$, $P_\Delta K_1$ is a Hilbert--Schmidt operator in $\mathcal H$. Hence, $K_1$ is an integral operator and its integral kernel $K_1(x,y)$ satisfies $\int_{\Delta\times X}|K_1(x,y)|^2\sigma(dx)\sigma(dy)<\infty$ for each $\Delta\in\mathcal B_0(X)$. Thus, we may assume that, for each $x\in X$,   $K_1(x,\cdot)\in\mathcal H$. 

Now, assume additionally that the integral kernel $K_1(x,y)$ is real-valued, i.e., the operator $K_1$ (equivalently the operator $K$) maps the  space  of 
real-valued functions from $\mathcal H$ into itself.  Set $\mathcal G=\mathcal H$, let $\mathcal I$ be the complex conjugation in $\mathcal H$, and define $L_1(x)=L_2(x)=K_1(x,\cdot)\in\mathcal H$ for $x\in X$. Then assumptions \eqref{xzra4yrde4} are trivially satisfied. In this case,
$$\mathcal K_1(x,y)=\mathcal K_2(x,y)=(K_1(x,\cdot),K_1(y,\cdot))_{\mathcal H}=K(x,y)=\overline{K(x,y)},$$
where $K(x,y)$ is the  integral kernel of the operator $K$. The correlation functions of the corresponding point process $\mu$ can be written in the form
$$k^{(n)}(x_1,\dots,x_n)=\operatorname{det}_2 [ K(x_i,x_j)]_{i,j=1,\dots,n}\,,$$
where $\operatorname{det}_2$ the the 2-determinant: for a matrix $B=[b_{ij}]_{i,j=1,\dots,n}$, 
\begin{equation*}
\operatorname{det}_2(B)=\sum_{\pi\in \mathfrak S_n}\prod_{i=1}^n2^{n-C(\pi)}\,b_{i\,\pi(i)}.\end{equation*}
Here $C(\pi)$ denotes the number of cycles in the permutation $\pi$. Such a point process $\mu$ is called $2$-permanental.


\begin{thebibliography}{99}


\bibitem{AL2}  Alshehri, M.G.A., Lytvynov, E.:
Hafnian point processes and quasi-free states on the CCR algebra. 
{\it Infin. Dimens. Anal. Quantum Probab. Relat. Top.} {\bf 25} (2022),  Paper No.~2250002, 25 pp.


\bibitem{AL1} Alshehri, M.G.A., Lytvynov, E.: Particle-hole transformation in the continuum and determinantal point processes. {\it Comm. Math. Phys.} {\bf 403} (2023),  627--659.

\bibitem{ArWoods} Araki, H., Woods, E.: Representations of the C.C.R. for a nonrelativistic infinite free Bose gas, {\it J. Math.\ Phys.}\ {\bf 4} (1963), 637--662.


\bibitem{AW}  Araki, H., Wyss, W.: Representations of canonical
anticommutation relation. {\it Helv.\ Phys.\ Acta.}\ {\bf 37} (1964), 136--159. 


\bibitem{ber_1984} Berezanskii, Yu.M.: The projection spectral theorem. {\it Russian Math. Surveys} {\bf 39} (1984), 1--62


\bibitem{ber_1985} Berezanskii, Yu.M.: On the projection spectral theorem. {\it Ukrainian Math. J.} {\bf 37} (1985), 124--130.

\bibitem{ber_1978}
Berezanskii, Yu.M.:  Selfadjoint operators in spaces of functions of infinitely many variables. Naukova Dumka, Kiev, 1978 (in Russian), English translation: American Mathematical Society, Providence, RI, 1986.

\bibitem{Ber} Berezansky, Y.M.: Poisson measure as the spectral measure of Jacobi field. {\it Infin. Dimens. Anal. Quantum Probab. Relat. Top.} {\bf 3} (2000), 121--139.


\bibitem{BK} Berezansky, Y.M., Kondratiev, Y.G.: Spectral methods in infinite-dimensional analysis. Naukova Dumka, Kyiv, 1988 (in Russian). English translation: Kluwer Academic Publishers, Dordrecht, 1995. 

\bibitem{BKKL} Berezansky, Y.M., Kondratiev, Y.G., Kuna, T., Lytvynov, E.: 
On a spectral representation for correlation measures in configuration space analysis.
{\it Methods Funct. Anal. Topology} {\bf 5} (1999), no.~4, 87--100.

 \bibitem{BUS} Berezansky, Y.M.,  Sheftel, Z.G., Us, G.F.: Functional Analysis, Vol.~1, Birkh\"auser, Basel, 1996.

\bibitem{BR}  Bratteli, O.,   Robinson, D.W.: Operator algebras and quantum statistical mechanics. Vol.~2. Equilibrium states. Models in quantum statistical mechanics. Second edition. Springer, Berlin, 1997.

\bibitem{KK} Kondratiev, Y.G., Kuna, T.: Harmonic analysis on configuration space. I. General theory. {\it Infin. Dimens. Anal. Quantum Probab. Relat. Top.} {\bf 5} (2002),  201--233.

\bibitem{Lenard}   Lenard, A.: Correlation functions and the uniqueness of the state in classical statistical mechanics, {\it Commun. Math. Phys.} {\bf 30} (1973), 35--44. 

\bibitem{L2} Lytvynov, E.:  Fermion and boson random point processes as particle distributions of infinite free Fermi and Bose gases of finite density. {\it Rev. Math. Phys.} {\bf 14} (2002), 1073--1098. 


\bibitem{L1} Lytvynov, E.: The projection spectral theorem and Jacobi fields. 
{\it Methods Funct. Anal. Topology} {\bf 21} (2015),  188--198.

\bibitem{LM} Lytvynov, E., Mei, L.: On the correlation measure of a family of commuting Hermitian operators with applications to particle densities of the quasi-free representations of the CAR and CCR,
{\it J. Funct. Anal.}  {\bf 245} (2007),  62--88. 

\bibitem{Parth} Parthasarathy, K.R.: An introduction to quantum stochastic calculus.  Birkh\"auser, Basel, 1992.



\end{thebibliography}
\end{document}